\documentclass[12pt]{article}
\usepackage{amssym}

\newcommand{\aalpha}{\overline{\alpha}}
\newcommand{\bbeta}{\overline{\beta}}
\newcommand{\ggamma}{\overline{\gamma}}
\newcommand{\qq}{\overline{q}}
\newcommand{\e}{\mbox{$\cal{E}$}}
\newcommand{\ee}{\overline{\e}}
\newcommand{\xb}{\mbox{\boldmath $x$}}
\newcommand{\pb}{\mbox{\boldmath $p$}}\def\case#1#2{{\textstyle{#1\over #2}}}
\newcommand{\te}{\tilde{e}}
\newcommand{\tepsilon}{\tilde{\epsilon}}

\title{
More on a SUSYQM approach to the harmonic oscillator with nonzero minimal uncertainties
in position and/or momentum}
\author{C Quesne$^{\dagger}$ and  V M Tkachuk$^{\ddagger}$\\
$^{\dagger}$ {\small Physique Nucl\'eaire Th\'eorique et Physique
Math\'ematique,  Universit\'e Libre de Bruxelles,} \\ 
{\small Campus de la Plaine CP229, Boulevard~du Triomphe, B-1050 Brussels,
Belgium}\\ 
$^{\ddagger}$ {\small Ivan Franko Lviv National University, Chair of Theoretical
Physics,}\\
{\small 12, Drahomanov Street, Lviv UA-79005, Ukraine}\\
{\small E-mail: cquesne@ulb.ac.be and tkachuk@ktf.franko.lviv.ua}}
\date{ }
\begin{document}
\baselineskip=20pt plus 1pt minus 1pt
%%%%%%%%%%%%%%%%%%%%%%%%%%%%%%%%%%%%%%%%%%%%%%%%%%%%%%%%%%
\maketitle
\begin{abstract}
We continue our previous application of supersymmetric quantum mechanical methods to
eigenvalue problems in the context of some deformed canonical commutation relations
leading to nonzero minimal uncertainties in position and/or momentum. Here we
determine for the first time the spectrum and the eigenvectors of a one-dimensional
harmonic oscillator in the presence of a uniform electric field in terms of the deforming
parameters $\alpha$, $\beta$. We establish that whenever there is a nonzero minimal
uncertainty in momentum, i.e., for $\alpha \ne 0$, the correction to the harmonic
oscillator eigenvalues due to the electric field is level dependent. In the opposite case, i.e.,
for $\alpha = 0$, we recover the conventional quantum mechanical picture of an overall
energy-spectrum shift even when there is a nonzero minimum uncertainty in position,
i.e., for $\beta \ne 0$. Then we consider the problem of a $D$-dimensional harmonic
oscillator in the case of isotropic nonzero minimal uncertainties in the position
coordinates, depending on two parameters $\beta$, $\beta'$. We extend our methods
to deal with the corresponding radial equation in the momentum representation and
rederive in a simple way both the spectrum and the momentum radial wave functions
previously found by solving the differential equation. This opens the way to solving new
$D$-dimensional problems. 
\end{abstract}

\vspace{0.5cm}


\newpage
\section{Introduction}

During recent years, there has been much interest in studying quantum mechanical
problems under the assumption of a modified Heisenberg uncertainty relation leading to
nonzero minimal uncertainties in position and/or momentum. Such works are motivated
by several independent lines of investigations in string theory and quantum gravity, which
suggest the existence of a finite lower bound to the possible resolution of length
$\Delta x_0$ (see, e.g., \cite{gross, maggiore, witten}). Furthermore, the absence of
plane waves or momentum eigenvectors on generic curved spaces also hint at
a finite lower bound to the possible resolution of momentum $\Delta p_0$ (see, e.g.,
\cite{kempf94a}).\par
%
%--------------------------------------------------------------------------------------------------------
%
Such nonzero minimal uncertainties in position and momentum can be described in the
framework of small quadratic corrections to the canonical commutation
relations~\cite{kempf94b, kempf97, kempf01}. These corrections can also provide an
effective description of non-pointlike particles, such as quasiparticles and various collective
excitations in solids, or composite particles, such as nucleons and
nuclei~\cite{kempf97}.\par
%
%--------------------------------------------------------------------------------------
% 
The resolution of quantum mechanical problems with such deformed canonical
commutation relations has been mostly restricted to the case where there is only a
nonzero minimal uncertainty in position. Then one may indeed consider a deformed
Schr\"odinger equation in momentum representation and solve it using the technique of
differential equations. An exact solution to the one-dimensional harmonic oscillator
problem has been obtained in this way~\cite{kempf95}. This approach has been 
extended to $D$ dimensions~\cite{chang} and some ladder operators have been
constructed~\cite{dadic}. Some perturbative or partial results have also been
obtained for the hydrogen atom~\cite{brau, akhoury}.\par
%
%---------------------------------------------------------------------------------
%
The case where there are nonzero minimal uncertainties in both position and momentum
is much more involved because there is neither position nor momentum representation,
so that one has to resort to a generalized Fock space representation or, equivalently, to
the corresponding Bargmann representation~\cite{kempf94b, kempf93}. In this context,
we have recently solved~\cite{cq03a} the eigenvalue problem for the one-dimensional
harmonic oscillator in a purely algebraic way by availing ourselves of an extension of
supersymmetric quantum mechanical (SUSYQM) and shape-invariance techniques. The
SUSYQM formalism~\cite{cooper, junker} supplemented with shape invariance under
parameter translation~\cite{gendenshtein, dabrowska} is known to be a reformulation of
the factorization method developed by Schr\"odinger~\cite{schrodinger} and by Infeld and
Hull~\cite{infeld} (for a comparison between these two methods and corresponding
references see \cite{junker, carinena}). The procedure that we have used relies on a
generalized type of shape invariance, namely that connected with parameter
scaling~\cite{spiridonov92, khare, lutzenko, loutsenko}.\par
%
%------------------------------------------------------------------------------------------------
%
The purpose of the present paper is to further illustrate the power of SUSYQM techniques
in solving eigenvalue problems corresponding to deformed canonical commutation
relations.\par
%
%----------------------------------------------------------------------------------------------
%
{}First, we will consider the case of a one-dimensional harmonic oscillator with nonzero
minimal uncertainties in position and momentum in the presence of a uniform electric
field. Contrary to the solution of the corresponding problem in conventional quantum
mechanics, which can be obtained from the solution without electric field by a simple
coordinate shift, that of the present problem is more involved and will reveal some new
features.\par
%
%---------------------------------------------------------------------------------------------
%
Next, we will show how to extend our method to higher-dimensional problems by
providing an alternative and simple solution to the eigenvalue problem for the
$D$-dimensional harmonic oscillator with isotropic nonzero minimal uncertainties in the
position coordinates, which was previously dealt with by the differential equation
technique~\cite{chang}.\par
%
%---------------------------------------------------------------------------------------------
%
Our paper is organized as follows. The one-dimensional harmonic oscillator in a uniform
electric field is considered in section~2. Section~3 deals with the $D$-dimensional
harmonic oscillator. Finally, section~4 contains the conclusion.\par
%
%=========================================================
%
\section{One-dimensional harmonic oscillator in a uniform electric field}

Let us consider the case of a particle of mass $m$ and charge $\qq$ in a
harmonic potential $\frac{1}{2} m \omega^2 x^2$ and a uniform electric field $\ee$
parallel to the $x$-axis. It is described by the Hamiltonian  
\begin{equation}
  H = \frac{p^2}{2m} + \frac{1}{2} m \omega^2 x^2 - \qq \ee x  \label{eq:H}
\end{equation}
and the corresponding eigenvalue problem reads
\begin{equation}
  H |\psi_n\rangle = E_n |\psi_n\rangle \qquad n = 0, 1, 2, \ldots . 
\end{equation}
Here $x$ and $p$ are assumed to satisfy the deformed canonical commutation relation
\begin{equation}
  [x, p] = {\rm i} \hbar (1 + \aalpha x^2 + \bbeta p^2)  \label{eq:com}  
\end{equation}
where $\aalpha \ge 0$, $\bbeta \ge 0$, and $\aalpha \bbeta < \hbar^{-2}$, so that the
minimal uncertainties in position and momentum are given by $\Delta x_0 = \hbar
\sqrt{\bbeta/(1 - \hbar^2 \aalpha \bbeta)}$ and $\Delta p_0 = \hbar \sqrt{\aalpha/(1 -
\hbar^2 \aalpha \bbeta)}$, respectively~\cite{kempf94b}.\par
%
%-------------------------------------------------------------------------------------------------------
% 
In terms of dimensionless operators, $X = x/a$, $P = pa/\hbar$, $h = H/(\hbar
\omega)$ and dimensionless parameters $\alpha = \aalpha a^2$ and $\beta = \bbeta
\hbar^2/a^2$, $\e = \qq \ee a/(\hbar \omega)$, where $a =
\sqrt{\hbar/(m\omega)}$, equations (\ref{eq:H})~-- (\ref{eq:com}) can be rewritten as
\begin{equation}
  h = \frac{1}{2} (P^2 + X^2) - \e X  \label{eq:h}
\end{equation}
\begin{equation}
  h |\psi_n\rangle = e_n |\psi_n\rangle \qquad e_n \equiv E_n/(\hbar \omega) \qquad n
  = 0, 1, 2, \ldots 
\end{equation}
\begin{equation}
  [X, P] = {\rm i}(1+ \alpha X^2 + \beta P^2)  \label{eq:com-bis}
\end{equation}
with $\alpha \ge 0$, $\beta \ge 0$, and $\alpha \beta < 1$.\par
%
%-------------------------------------------------------------------------------
%
\subsection{Energy spectrum}

At this stage, it is worth noting that going to transformed operators $X' = X - \e$ and
$P' = P$, as in conventional quantum mechanics, would convert $h$ into a shifted
harmonic oscillator Hamiltonian
\begin{equation}
  h = \frac{1}{2} (P^{\prime 2} + X^{\prime 2}) - \frac{1}{2} \e^2  \label{eq:transf-h}  
\end{equation}
but, at the same time, change the commutation relation (\ref{eq:com-bis}) into
\begin{equation}
  [X', P'] = {\rm i}(1+ \alpha X^{\prime 2} + \beta P^{\prime 2} + 2 \alpha \e X' +
  \alpha \e^2).  \label{eq:transf-com} 
\end{equation}
Hence, except for $\alpha=0$, the energy spectrum of $h$ cannot be deduced from that
for $\e=0$ obtained in our previous paper~\cite{cq03a} (henceforth referred to as I and
whose equations will be subsequently quoted by their number preceded by I).\par
%
%----------------------------------------------------------------------------------------
%
Let us instead try to factorize $h$ as
\begin{equation}
  h = B^+(g,s,r) B^-(g,s,r) + \epsilon_0  \label{eq:fact-h}
\end{equation}
where \begin{equation}
  B^{\pm}(g,s,r) = \frac{1}{\sqrt{2}} (s X \mp {\rm i} g P + r)  \label{eq:B+/-}
\end{equation}
and $\epsilon_0$ is the factorization energy. In (\ref{eq:B+/-}), $g$, $s$ and $r$ are
assumed to be three constants depending on the parameters $\alpha$, $\beta$, $\e$ of
the problem. The first two, $g$ and $s$, are chosen positive and going to 1 in the limit
$\alpha$, $\beta \to 0$, while $r$ is taken as going to $-\e$ in the same limit. With
this choice, the operators $B^+(g,s,r)$ and $B^-(g,s,r)$ will be counterparts of the
creation and annihilation operators for the shifted harmonic oscillator occurring in
conventional quantum mechanics.\par
%
%-----------------------------------------------------------------------------------------
%
On inserting (\ref{eq:B+/-}) in (\ref{eq:fact-h}) and comparing the resulting expression
with (\ref{eq:h}), we get four conditions, of which the first two are independent of $\e$
and coincide with equations (I2.7) and (I2.8). They fix the values of $g$ and $s$, 
\begin{equation}
  g = sk \qquad s = \frac{1}{\sqrt{1 - \alpha k}} \qquad k \equiv \case{1}{2}(\beta -
  \alpha) + \sqrt{1 + \case{1}{4}(\beta - \alpha)^2}  \label{eq:k}
\end{equation}
in terms of $\alpha$ and $\beta$. The remaining two conditions read
\begin{eqnarray}
  rs & = & - \e \label{eq:fact-cond3} \\
  \epsilon_0 & = & \frac{1}{2} (gs - r^2). \label{eq:fact-cond4}
\end{eqnarray}
Equation (\ref{eq:fact-cond3}) provides the value of $r$,
\begin{equation}
  r = - \e \sqrt{1 - \alpha k}
\end{equation}
while equation (\ref{eq:fact-cond4}) leads to the value of the factorization energy in
terms of the three parameters $\alpha$, $\beta$, $\e$.\par
%
%-------------------------------------------------------------------------------------------------
% 
Having proved that the Hamiltonian $h$ is factorizable, let us now consider a hierarchy of
Hamiltonians
\begin{equation}
  h_i = B^+(g_i, s_i,r_i) B^-(g_i, s_i,r_i) + \sum_{j=0}^i \epsilon_j \qquad  i=0, 1, 2,
  \ldots \label{eq:hierarchy}
\end{equation}
whose first member $h_0$ coincides with $h$. Here $g_i$, $s_i$, $\epsilon_i$, $r_i$
are some parameters, the first three being assumed positive, and $g_0 = g$, $s_0 = s$,
$r_0 = r$.\par
%
%---------------------------------------------------------------------------------------------
%
Proceeding as in I, let us impose the shape invariance condition
\begin{equation}
  B^-(g_i, s_i, r_i) B^+(g_i, s_i, r_i) = B^+(g_{i+1}, s_{i+1}, r_{i+1}) B^-(g_{i+1},
  s_{i+1}, r_{i+1}) + \epsilon_{i+1}  \label{eq:SI}
\end{equation}
where $i=0$, 1, 2,~\ldots. It is equivalent to a set of four relations, of which the first
two are again independent of $\e$ and coincide with equations (I2.15) and (I2.16),
respectively. In I, the latter have been solved by introducing some new combinations of
parameters
\begin{eqnarray}
  u_i & = & g_i + \gamma s_i \qquad v_i = g_i - \gamma s_i \qquad \gamma \equiv
        \sqrt{\frac{\beta}{\alpha}}  \label{eq:ui} \\
  d_i & = & u_i v_i \qquad t_i = \frac{v_i}{u_i}
\end{eqnarray}
thereby leading to 
\begin{equation}
  d_i = d \qquad t_i = q^{-i} t \qquad \mbox{\rm or} \qquad u_i = q^{i/2} u \qquad
  v_i = q^{-i/2} v  \label{eq:di}
\end{equation}
where
\begin{equation}
  d \equiv uv \qquad t \equiv \frac{v}{u} \qquad q \equiv \frac{1 +
  \sqrt{\alpha\beta}}{1 - \sqrt{\alpha\beta}}.  \label{eq:d}
\end{equation}
\par
%
%-----------------------------------------------------------------------------------
%
The remaining two relations read
\begin{eqnarray}
  r_{i+1} s_{i+1} & = & r_i s_i  \label{eq:SI-cond3} \\
  \epsilon_{i+1} & = & \frac{1}{2}(g_i s_i + g_{i+1} s_{i+1} + r_i^2 - r_{i+1}^2).
      \label{eq:SI-cond4}
\end{eqnarray}
On taking equations (\ref{eq:fact-cond3}), (\ref{eq:ui}), (\ref{eq:di}), and (\ref{eq:d})
into account, we obtain from (\ref{eq:SI-cond3}) the solution for $r_i$,
\begin{equation}
  r_i = - \frac{2\gamma\e}{u} q^{-i/2} \left(1 - \frac{t}{q^i}\right)^{-1}.
\end{equation}
\par
%
%--------------------------------------------------------------------------------------------
% 
{}Furthermore, equation (\ref{eq:SI-cond4}) leads to the eigenvalues $e_n$ of $h$,
\begin{equation}
  e_n(q,t,\e) = \sum_{i=0}^n \epsilon_i = \sum_{i=0}^{n-1} g_i s_i + \frac{1}{2} g_n
         s_n - \frac{1}{2} r_n^2 = e_n(q,t,0) + \Delta e_n(q,t,\e)  \label{eq:spectrum}
\end{equation}
where
\begin{equation}
  e_n(q,t,0) = \frac{K^2}{q+1} \left\{\left(1 - \frac{t^2}{q^{n-1}}\right) [n]_q + 
         \frac{1}{2} \left(q^n - \frac{t^2}{q^n}\right)\right\} \qquad K \equiv u
         \sqrt{\frac{q+1}{4\gamma}} \qquad [n]_q \equiv \frac{q^n-1}{q-1} 
         \label{eq:spectrum-0} 
\end{equation}
are the eigenvalues in the absence of electric field, given in (I2.30), and
\begin{equation}
  \Delta e_n(q,t,\e) = - \frac{2 \gamma^2 \e^2}{u^2} q^{-n} \left(1 -
  \frac{t}{q^n}\right)^{-2}  \label{eq:correction}
\end{equation}
are the corrections due to the electric field. The latter can be rewritten as
\begin{equation}
  \Delta e_n(q,t,z) = - \frac{1}{2} K^2 z^2 (1-t)^2 q^{-n} \left(1 -
  \frac{t}{q^n}\right)^{-2}  \label{eq:correction-bis}
\end{equation}
in terms of a new parameter proportional to $\e$,
\begin{equation}
  z \equiv 4 \gamma \sqrt{\frac{\gamma}{q+1}} \frac{\e}{u^2(1-t)}
\end{equation}
which will subsequently prove convenient (see equations (\ref{eq:zi-spe}) and
(\ref{eq:zi})).\par
%
%-------------------------------------------------------------------------------------
%
In contrast with conventional quantum mechanics, the additional negative contributions
$\Delta e_n(q,t,z)$ are $n$ dependent for generic $\alpha$ and $\beta$ values. Since
$q>1$, their absolute value decreases from $\frac{1}{2} K^2 z^2$ to 0 when $n$ goes from
0 to $\infty$. As it can be inferred from the definition (\ref{eq:h}) of $h$, where the electric
field breaks the symmetry under exchange of $X$ and $P$, $\Delta e_n(q,t,z)$ is not invariant
under exchange of $\alpha$ and $\beta$ as it is the case for $e_n(q,t,0)$. This will lead to
different limiting cases in the next subsection.\par
%
%-------------------------------------------------------------------------------------------
%
The corrections to the excitation energies in the absence of electric field
\begin{equation}
  e_n(q,t,0) - e_0(q,t,0) = \frac{1}{2} K^2 \left(1 - \frac{t^2}{q^n}\right) [n]_q 
\end{equation}
are given by
\begin{equation}
  \Delta e_n(q,t,z) - \Delta e_0(q,t,z) = \frac{1}{2} (q-1) K^2 z^2 q^{-n} \left(1 -
  \frac{t^2}{q^n}\right) \left(1 - \frac{t}{q^n}\right)^{-2} [n]_q.
\end{equation}
\par
%
%----------------------------------------------------------------------------------------
%
On taking the results for $g_i$, $s_i$, $r_i$, and $\epsilon_i$ into account, the
Hamiltonians (\ref{eq:hierarchy}) of the SUSYQM hierarchy can be written as
\begin{equation}
  h_i = \frac{1}{2}(a_i P^2 + b_i X^2) - \e X + c_i \qquad i=0, 1, 2, \ldots
\end{equation}
where $a_i$, $b_i$, $c_i$ are constants independent of $\e$ and given by equation
(I2.36). Going back to variables with dimensions, we get
\begin{equation}
  H_i \equiv \hbar \omega h_i = \frac{p^2}{2m_i} + \frac{1}{2} m_i \omega_i^2 x^2
  - \qq \ee x+ c_i \hbar \omega \qquad m_i = \frac{m}{a_i} \qquad \omega_i =
\sqrt{a_i b_i}\, \omega.  
\end{equation}
Note that contrary to the mass and the frequency, the electric field is the same for all
the partners.
\par
%
%++++++++++++++++++++++++++++++++++++++++++++++++++++++
%
\subsection{Some special cases}

\subsubsection{\boldmath Limit $\alpha \to 0$}

{}For small $\alpha$ values, we obtain 
\begin{equation}
  q^{-n} \simeq 1 - 2n \sqrt{\alpha\beta} + O(\alpha) \qquad \frac{u}{\gamma} \left(1 -
  \frac{t}{q^n}\right) \simeq 2s \left(1 - n \sqrt{\alpha\beta}\right) + O(\alpha).
\end{equation}
Inserting such results in (\ref{eq:correction}) transforms the latter into
\begin{equation}
  \Delta e_n \simeq - \frac{\e^2}{2s^2} + O(\alpha) \qquad \mbox{\rm where} \qquad
  s \simeq 1 + O(\alpha).
\end{equation}
\par
%
%--------------------------------------------------------------------
%
In the limit $\alpha \to 0$, we therefore get
\begin{equation}
  \Delta e_n(\e) = - \frac{1}{2} \e^2
\end{equation}
which is independent  of $n$ and of $\beta$ and is actually the same as in conventional
quantum mechanics. We conclude that the $n$-dependent corrections to an exponential
spectrum obtained in the general case reduce to $n$-independent  corrections to a
quadratic spectrum in the special case $\alpha=0$. Such a result would also follow
from considering equations (\ref{eq:transf-h}) and (\ref{eq:transf-com}) for
$\alpha=0$.\par
%
%xxxxxxxxxxxxxxxxxxxxxxxxxxxxxxxxxxxxxxxxxxxxxxxxxxxxxxxxxxxx
%
\subsubsection{\boldmath Limit $\beta \to 0$}

{}For small $\beta$ values, we obtain 
\begin{equation}
  q^{-n} \simeq 1 + 2n \sqrt{\alpha\beta} + O(\beta) \qquad \frac{u}{\gamma} \left(1 -
  \frac{t}{q^n}\right) \simeq 2(s + ng\alpha) + O\left(\sqrt{\beta}\right)
\end{equation}
so that equation (\ref{eq:correction}) now becomes
\begin{equation}
  \Delta e_n \simeq - \frac{\e^2}{2(s + ng\alpha)^2} + O\left(\sqrt{\beta}\right) 
\end{equation}
where
\begin{equation}
  g \simeq 1 + O(\beta) \qquad s \simeq \frac{1}{2} \alpha + \sqrt{1 + \frac{1}{4}
  \alpha^2} + O (\beta).
\end{equation}
\par
%
%---------------------------------------------------------------
%
In the limit $\beta \to 0$, we therefore get corrections
\begin{equation}
  \Delta e_n(\alpha, \e) = - \frac{1}{2} \e^2 \left[\left(n + \frac{1}{2}\right) \alpha +
  \sqrt{1 + \frac{1}{4} \alpha^2}\right]^{-2}  \label{eq:correction-spe}
\end{equation}
to a quadratic spectrum $e_n(\alpha)$. The latter can be obtained from equation (I3.6)
by substituting $\alpha$ for $\beta$. We conclude that the essential feature of the
general case, namely the $n$ dependence of the correction terms, is already present
when there is only a nonzero minimal uncertainty in momentum.\par
%
%--------------------------------------------------------
%
The result (\ref{eq:correction-spe}) may also be derived directly from SUSYQM and
shape invariance without resorting to a limiting procedure. Going back to the
factorization and shape invariance conditions given in section~2.1 and setting
$\beta=0$ therein, we are only left with two parameters $s_i$ and $r_i$, satisfying the
conditions $s_{i+1} (s_{i+1} - \alpha) = s_i (s_i + \alpha)$, $r_{i+1} = r_i s_i/s_{i+1}$,
since $g_i = g = 1$. This leads to $s_i = s + i \alpha$ and $r_i = - \e/s_i$, from which
equation (\ref{eq:correction-spe}) can be easily obtained. The fact that $r_i$ is not
independent of $i$, contrary to what happens for $\alpha=0$, is clearly responsible for
the $n$ dependence of $\Delta e_n(\alpha, \e)$ shown in (\ref{eq:correction-spe}).\par
%
%xxxxxxxxxxxxxxxxxxxxxxxxxxxxxxxxxxxxxxxxxxxxxxxxxxxxxxxxx
%
\subsubsection{\boldmath Case $\alpha = \beta \ne 0$}

In I, we showed that whenever the two dimensionless deforming parameters $\alpha$
and $\beta$ are equal (which means that there is a specific relation $\aalpha = m^2
\omega^2 \bbeta$ between the original parameters $\aalpha$, $\bbeta$ of equation
(\ref{eq:com})), the harmonic oscillator Hamiltonian $\frac{1}{2} (P^2 + X^2)$ reduces to
a $q$-deformed harmonic oscillator one
\begin{equation}
  h_{\rm osc} = \case{1}{4} (q+1) \{b, b^+\}  \label{eq:h-osc}  
\end{equation}
where
\begin{equation}
  b^+ = \frac{1}{\sqrt{q+1}} (X - {\rm i}P) \qquad b = \frac{1}{\sqrt{q+1}} (X + {\rm
  i}P)  \label{eq:b+/-}
\end{equation}
satisfy the relation
\begin{equation}
  b b^+ - q b^+ b = I  \label{eq:q-com}
\end{equation}
with $q = (1+\alpha)/(1-\alpha)$.\par
%
%------------------------------------------------------------
%
In terms of the operators (\ref{eq:h-osc}) and (\ref{eq:b+/-}), the Hamiltonian
(\ref{eq:h}) can be rewritten as
\begin{equation}
  h = h_{\rm osc} - \frac{1}{2} \sqrt{q+1}\, \e (b^+ + b).
\end{equation}
Since for $\alpha = \beta$, we get $\gamma=1$, $t=0$ and $u = 2/\sqrt{1-\alpha} =
\sqrt{2(q+1)}$, it follows from equation (\ref{eq:correction}) or
(\ref{eq:correction-bis}) that the correction term to the eigenvalues (I3.11) of $h_{\rm
osc}$ reads
\begin{equation}
  \Delta e_n(q, \e) = - \frac{\e^2}{q+1} q^{-n}
\end{equation}
or
\begin{equation}
  \Delta e_n(q,z) = - \frac{1}{2} K^2 z^2 q^{-n} \qquad K = \frac{1}{\sqrt{2}}(q+1) \qquad
  z =2 \e (q+1)^{-3/2}
\end{equation}
which is again $n$ dependent.\par
%
%+++++++++++++++++++++++++++++++++++++++++++++++++++++++++++++
%
\subsection{Hamiltonian eigenvectors}

Before going to the general case $0 \ne \alpha \ne \beta \ne 0$, it is worth considering the
special case $\alpha = \beta \ne 0$, for which the Hamiltonian eigenvalues and eigenvectors
depend only on two parameters instead of three.\par
%
%xxxxxxxxxxxxxxxxxxxxxxxxxxxxxxxxxxxxxxxxxxxxxxxxxxxxxxxxxxxxxx
%
\subsubsection{\boldmath Case $\alpha = \beta \ne 0$}

In terms of the $q$-deformed boson creation and annihilation operators $b^+$, $b$, defined
in equation (\ref{eq:b+/-}), the operators $B^{\pm}(g_i, s_i, r_i)$ corresponding to the
Hamiltonian (\ref{eq:hierarchy}) of the SUSYQM hierarchy, assume the simple form
\begin{equation}
  B^+(q, z_i) = \frac{1}{\sqrt{2}} K q^{i/2} (b^+ - z_i) \qquad B^-(q, z_i) =
  \frac{1}{\sqrt{2}} K q^{i/2} (b - z_i)  \label{eq:B-spe} 
\end{equation}
with
\begin{equation}
  z_i \equiv z q^{-i}.  \label{eq:zi-spe}
\end{equation}
In the ($q$-deformed) Bargmann representation of the operators $b^+$, $b$, associated with
the corresponding $q$-deformed coherent states~\cite{spiridonov95, cq03b}, they are
represented by
\begin{equation}
  {\cal B}^+(q, z_i) = \frac{1}{\sqrt{2}} K q^{i/2} (\xi - z_i) \qquad {\cal B}^-(q, z_i) =
  \frac{1}{\sqrt{2}} K q^{i/2} ({\cal D}_q - z_i) 
\end{equation}
where $\xi$ is a complex variable and ${\cal D}_q$ is the $q$-differential operator defined by
${\cal D}_q \psi(\xi) = [\psi(q\xi) - \psi(\xi)]/[(q-1)\xi]$.\par
%
%---------------------------------------------------------------------------------------------------
%
Here it should be stressed that although the electric field remains constant in the Hamiltonians
$h_i$ of the SUSYQM hierarchy, the corresponding parameter $z_i$ in the first-order
operators $B^{\pm}(q,z_i)$, involved in the factorization of $h_i$, is $i$ dependent, as shown
in (\ref{eq:zi-spe}).\par
%
%---------------------------------------------------------------------------------------
%
The ground state of $h$ is represented by a function $\psi_0(q,z;\xi)$, which is a normalized
solution of the equation 
\begin{equation}
  {\cal B}^-(q,z) \psi_0(q,z;\xi) = 0 \qquad \mbox{\rm or} \qquad {\cal D}_q
  \psi_0(q,z;\xi) = z \psi_0(q,z;\xi). 
\end{equation}
It is therefore the Bargmann representation of a $q$-deformed CS $|z\rangle_q$ with real $z$
and it is given by~\cite{cq03b}
\begin{equation}
  \psi_0(q,z;\xi) = {\cal N}_0(q,z) E_q(z\xi) \qquad {\cal N}_0(q,z) = [E_q(|z|^2)]^{-1/2} 
  \label{eq:gs-spe}
\end{equation}
where $E_q(\xi) = \sum_{n=0}^{\infty} \xi^n/[n]_q!$ is a $q$-exponential.\par
%
%-----------------------------------------------------------------------------------------
%
The normalized excited state Bargmann wave functions can be determined recursively through
the equations
\begin{eqnarray}
  \psi_{n+1}(q,z;\xi) & = & [e_{n+1}(q,z) - e_0(q,z)]^{-1/2} {\cal B}^+(q,z) \psi_n(q,z_1;
       \xi) \nonumber \\
  & = & \left\{[n+1]_q \left(1+ \frac{(q-1) z^2}{q^{n+1}}\right)\right\}^{-1/2} (\xi - z)
       \psi_n\left(q, \frac{z}{q}; \xi\right) 
\end{eqnarray}
where $n=0$, 1, 2,~\ldots. It is straightforward to show that they are given by 
\begin{equation}
  \psi_n(q,z;\xi) = {\cal N}_n(q,z) \prod_{k=0}^{n-1} \left(\xi - \frac{z}{q^k}\right)
   E_q\left(\frac{z}{q^n} \xi\right)  \label{eq:es-spe}
\end{equation}
where
\begin{equation}
  {\cal N}_n(q,z) = \left\{[n]_q! \Bigl(q^{-2n+1}(1-q)z^2; q\Bigr)_n\right\}^{-1/2}
  {\cal N}_0\left(q,\frac{z}{q^n}\right)  \label{eq:norm-n}
\end{equation}
and $n=1$, 2,~\ldots. In (\ref{eq:norm-n}), the symbol $(a;q)_n$ is defined as in equation
(IA.5).\par
%
%--------------------------------------------------------------------------------------
%
{}For $z \to 0$, the wave functions $\psi_n(q,z;\xi)$ reduce to the Bargmann representation
$\varphi_n(q;\xi) = \xi^n/\sqrt{[n]_q!}$ of the $n$-$q$-boson states $|n\rangle_q =
(b^+)^n/\sqrt{[n]_q!}\, |0\rangle_q$, which are the eigenvectors of $h_{\rm osc}$.\par
%
%xxxxxxxxxxxxxxxxxxxxxxxxxxxxxxxxxxxxxxxxxxxxxxxxxxxxxxxxxxxx
%
\subsubsection{\boldmath General case $0 \ne \alpha \ne \beta \ne 0$}

In the general case of unequal parameters $\alpha$ and $\beta$, $q$-boson creation and
annihilation operators satisfying equation (\ref{eq:q-com}) are defined by
\begin{equation}
  b^+ = \frac{1}{\sqrt{q+1}} \left(\frac{1}{\sqrt{\gamma}} X - {\rm i} \sqrt{\gamma}
  P\right) \qquad b = \frac{1}{\sqrt{q+1}} \left(\frac{1}{\sqrt{\gamma}} X + {\rm i}
  \sqrt{\gamma} P\right).
\end{equation}
The operators $B^{\pm}(g_i, s_i, r_i)$, corresponding to $h_i$, can be written in terms of
them as 
\begin{equation}
  B^+(q, t_i, z_i) = \frac{1}{\sqrt{2}} K q^{i/2} (b^+ - t_i b - z_i) \qquad B^-(q, t_i, z_i) =
  \frac{1}{\sqrt{2}} K q^{i/2} (b - t_i b^+ - z_i)  \label{eq:B-b}  
\end{equation}
and are represented by
\begin{equation}
  {\cal B}^+(q, t_i, z_i) = \frac{1}{\sqrt{2}} K q^{i/2} (\xi - t_i {\cal D}_q - z_i) \qquad 
  {\cal B}^-(q, t_i, z_i) = \frac{1}{\sqrt{2}} K q^{i/2} ({\cal D}_q - t_i \xi - z_i) 
  \label{eq:B-Barg}  
\end{equation}
in Bargmann representation. In equations (\ref{eq:B-b}) and (\ref{eq:B-Barg}), $z_i$ is given
by
\begin{equation}
  z_i = - \frac{r_i}{K q^{i/2}} = z q^{-i} (1-t) \left(1 - \frac{t}{q^i}\right)^{-1}  \label{eq:zi}
\end{equation}
and reduces to the value given in (\ref{eq:zi-spe}) for $\alpha = \beta$ or $t=0$.\par
%
%-----------------------------------------------------------------------------------------
%
The ground state Bargmann wave function now satisfies the first-order difference equation
\begin{equation}
  {\cal D}_q \psi_0(q,t,z;\xi) = (t\xi + z) \psi_0(q,t,z;\xi).  \label{eq:gs-eqn} 
\end{equation}
From the latter, it is clear that for large values of $|\xi|$, the behaviour of $\psi_0(q,t,z;\xi)$
is the same as that of $\psi_0(q,t;\xi)$, corresponding to $\e=0$. It therefore follows from I
that equation (\ref{eq:gs-eqn}) has a normalizable solution on the whole complex plane. By
using properties (IA.2) and (IA.8) of the $q$-exponential and the $q$-differential operator,
respectively, the latter can be written as
\begin{equation}
  \psi_0(q,t,z;\xi) = {\cal N}_0(q,t,z) E_q(\lambda \xi) E_q(\mu \xi)  \label{eq:gs} 
\end{equation}
where
\begin{equation}
  \lambda = \frac{1}{2} z + \Delta \qquad \mu = \frac{1}{2} z - \Delta \qquad \Delta =
  \sqrt{\frac{1}{4} z^2 - \frac{t}{q-1}}  \label{eq:lambda}  
\end{equation}
and ${\cal N}_0(q,t,z)$ is some normalization coefficient.\par
%
%-----------------------------------------------------------------------------------------
%
On expanding the two $q$-exponentials in (\ref{eq:gs}), $\psi_0(q,t,z;\xi)$ becomes
\begin{equation}
  \psi_0(q,t,z;\xi) = {\cal N}_0(q,t,z) \sum_{n=0}^{\infty} c_n(q,t,z) \varphi_n(q;\xi)
\end{equation}
where
\begin{eqnarray}
  c_n(q,t,z) & = & \frac{1}{\sqrt{[n]_q!}} \sum_{m=0}^n 
  \left[\begin{array}{c}        
      n \\ m
  \end{array}\right]_q 
  \lambda^m \mu^{n-m} = \frac{1}{\sqrt{[n]_q!}} \sum_{p=0}^{[n/2]} a_{n,p}(q) z^{n-2p}
  t^p \\
  a_{n,0}(q) & = & 1 \\
  a_{n,p}(q) & = & \sum_{r_1=2p-1}^{n-1} [r_1]_q \sum_{r_2=2p-3}^{r_1-2} [r_2]_q
       \cdots \sum_{r_i=2p-2i+1}^{r_{i-1}-2} [r_i]_q \cdots \sum_{r_p=1}^{r_{p-1}-2}
       [r_p]_q \nonumber  \\
  && p=1, 2, \ldots, [n/2]  \label{eq:anp} \\
  {\cal N}_0(q,t,z) & = & \left(\sum_{n=0}^{\infty} c_n^2(q,t,z)\right)^{-1/2}. 
\end{eqnarray}
Here $[n/2]$ denotes the largest integer contained in $n/2$ and $\left[\begin{array}{c}        
n \\ m \end{array}\right]_q = [n]_q!/([m]_q!\, [n-m]_q!)$ is a $q$-binomial coefficient. The
proof of equation (\ref{eq:anp}) is by induction over $p$.\par
%
%---------------------------------------------------------------------------------------------
%  
Note that $\Delta$, defined in (\ref{eq:lambda}), may be real or imaginary according to the
relative values of the parameters. For instance, for $\alpha = \beta$, we get  $t=0$ and
$\Delta = z/2$ so that $\lambda = z$ and $\mu = 0$ are real. In such a case, equation
(\ref{eq:gs}) reduces to equation (\ref{eq:gs-spe}). On the contrary, for $\e=0$, we get
$z=0$ and $\Delta = {\rm i} \sqrt{t/(q-1)}$ so that $\lambda = {\rm i} \sqrt{t/(q-1)}$ and
$\mu = \lambda^*$ are imaginary. Equation (\ref{eq:gs}) then becomes
\begin{equation}
  \psi_0(q,t;\xi) = {\cal N}_0(q,t) E_q\left({\rm i} \sqrt{\frac{t}{q-1}}\,\xi\right)
  E_q\left(-{\rm i} \sqrt{\frac{t}{q-1}}\,\xi\right) = {\cal N}_0(q,t)
  E_{q^2}\left(\frac{t}{q+1} \xi^2\right) \label{eq:gs-I}
\end{equation}
where in the last step we used a well-known property of the $q$-exponential~\cite{ubriaco,
cq03c}. Equation (\ref{eq:gs-I}) coincides with equation (I4.12).\par
%
%------------------------------------------------------------------------------------------
%
The normalized excited state Bargmann wave functions satisfy the recursion relation
\begin{eqnarray}
  \psi_{n+1}(q,t,z;\xi) & = & \left\{[n+1]_q \left(1 - \frac{t^2}{q^{n+1}}\right) \left[1+
        (q-1) z^2 q^{-n-1} \left(1 - \frac{t}{q^{n+1}}\right)^{-2}\right] \right\}^{-1/2}
        \nonumber \\
  && \mbox{} \times (\xi - t{\cal D}_q - z) \psi_n(q, t_1, z_1; \xi) \qquad n = 0, 1, 2, \ldots
        \label{eq:recursion}
\end{eqnarray}
where $t_1 = t/q$ and $z_1 = (z/q) (1-t) (1-t/q)^{-1}$.\par
%
%------------------------------------------------------------------------------------------
%
The solution of equation (\ref{eq:recursion}) can be written as
\begin{equation}
  \psi_n(q,t,z;\xi) = {\cal N}_n(q,t,z) P_n(q,t,z;\xi) E_q(\lambda_n \xi) E_q(\mu_n \xi) 
  \label{eq:excited}
\end{equation}
where $\lambda_n$ and $\mu_n$ are given by equation (\ref{eq:lambda}) with $t$ and $z$
replaced by $t_n$ and $z_n$, respectively, $P_n(q,t,z;\xi)$ is an $n$th-degree polynomial in
$\xi$, satisfying the relation
\begin{equation}
  P_{n+1}(q,t,z;\xi) = (\xi-z) P_n(q, t_1, z_1; \xi) - t (t_{n+1}\xi + z_{n+1})  P_n(q, t_1, z_1;
  q\xi) - t {\cal D}_q P_n(q, t_1, z_1; \xi) 
\end{equation}
with $P_0(q,t,z;\xi) \equiv 1$, and ${\cal N}_n(q,t,z)$ is a normalization coefficient fulfilling
the recursion relation
\begin{eqnarray}
  {\cal N}_{n+1}(q,t,z) & = & \left\{[n+1]_q \left(1 - \frac{t^2}{q^{n+1}}\right) \left[1+
        (q-1) z^2 q^{-n-1} \left(1 - \frac{t}{q^{n+1}}\right)^{-2}\right] \right\}^{-1/2}
        \nonumber \\
  && \mbox{} \times {\cal N}_n(q, t_1, z_1).
\end{eqnarray}
\par
%
%------------------------------------------------------------------------------------------------
%
It can be shown that for the first few $n$ values, the polynomials $P_n(q,t,z;\xi)$ are given
by
\begin{eqnarray}
  P_1(q,t,z;\xi) & = & \left(1 - \frac{t^2}{q}\right) \Biggl[\xi - z\left(1 -
       \frac{t}{q}\right)^{-1}\Biggr] \label{eq:P1}\\
  P_2(q,t,z;\xi) & = & \left(1 - \frac{t^2}{q^3}\right) \Biggl[\left(1 -
       \frac{t^2}{q}\right)\xi^2 - [2]_q \frac{z}{q}\left(1 - \frac{t^2}{q}\right) \left(1 -
       \frac{t}{q^2}\right)^{-1} \xi - t \nonumber \\
  && \mbox{} + \frac{z^2}{q} (1-t) \left(1 + \frac{t}{q}\right) \left(1 -
       \frac{t}{q^2}\right)^{-2}\Biggr] \label{eq:P2} \\
  P_3(q,t,z;\xi) & = & \left(1 - \frac{t^2}{q^5}\right) \left(1 - \frac{t^2}{q^3}\right)
       \Biggl\{\left(1 - \frac{t^2}{q}\right)\xi^3 - [3]_q \frac{z}{q^2}\left(1 -
       \frac{t^2}{q}\right) \left(1 - \frac{t}{q^3}\right)^{-1} \xi^2  \nonumber \\
  && \mbox{} + [3]_q \Biggl[- \frac{t}{q} + \frac{z^2}{q^3} (1-t) \left(1 + \frac{t}{q}\right)
       \left(1 - \frac{t}{q^3}\right)^{-2}\Biggr] \xi  \nonumber \\
  && \mbox{} + z \frac{t}{q^2} \Bigl[([2]_q + q) - ([2]_q + 1)t\Bigr] \left[\left(1 -
       \frac{t}{q}\right)\left(1 - \frac{t}{q^3}\right)\right]^{-1} \nonumber \\
  && \mbox{} - \frac{z^3}{q^3} (1-t)^2 \left(1 + \frac{t}{q^2}\right) \Biggl[\left(1 -
       \frac{t}{q}\right)\left(1 - \frac{t}{q^3}\right)^3\Biggr]^{-1}\Biggr\}. \label{eq:P3}
\end{eqnarray}
For $z=0$, they reduce to the corresponding polynomials obtained in (I4.26) -- (I4.28), while
for $t=0$, they give back the polynomials multiplying the $q$-exponential in equation
(\ref{eq:es-spe}), as it should be.\par
%
%xxxxxxxxxxxxxxxxxxxxxxxxxxxxxxxxxxxxxxxxxxxxxxxxxxxxxx
%
\subsection{Comparison with a previous work}

From equation (\ref{eq:fact-h}) and equation (\ref{eq:B-b}) corresponding to $i=0$, it
follows that the Hamiltonian $h$, defined in (\ref{eq:h}), is a Hermitian form bilinear in
the $q$-boson creation and annihilation operators $b^+$, $b$, satisfying equation
(\ref{eq:q-com}) with $q>1$. The spectrum of the most general abstract Hamiltonian of
such a type was investigated some years ago~\cite{loutsenko}. It is therefore interesting
to see how our results compare with those previously obtained.\par
%
%--------------------------------------------------------------
%
In \cite{loutsenko}, it has been shown that there exist two factorization schemes leading
to two possible branches of the discrete spectrum for the considered Hamiltonian. In our
case too, the existence of such schemes has been noticed: the parameter $k \equiv g/s$
indeed satisfies some quadratic equation, of which we have only kept the root with a
plus sign in front of the radical, as shown in (\ref{eq:k}) (see also~\cite{cq03a}). Our
motivation for eliminating the other root with a minus sign in front of the radical has
been purely physical. We indeed consider as previous authors working in the field of very
small quadratic corrections to the canonical commutation relations leading to nonzero
minimal uncertainties in position and momentum~\cite{kempf94a, kempf94b, kempf97,
kempf01, kempf95, chang, dadic, brau, akhoury, kempf93} that our theory should give back
the standard results when the deforming parameters $\alpha$, $\beta$ go to zero. The
other solution for $k$ appears to be connected with a so-called classically singular
representation (see e.g.~\cite{aizawa}), which is regarded in such a theory as
unphysical.\par
%
%---------------------------------------------------------------------
% 
{}Focusing now our attention  on the first factorization scheme of~\cite{loutsenko}, we
notice that our result for the energy spectrum, given in (\ref{eq:spectrum}),
(\ref{eq:spectrum-0}) and (\ref{eq:correction-bis}), can be retrieved from equation (20)
of~\cite{loutsenko} after identifying the parameters $\alpha_0$, $\beta_0$ and
$\gamma_0$ of equation (23) in the same reference with $K/\sqrt{2}$, $-Kt/\sqrt{2}$
and $-Kz/\sqrt{2}$, respectively.\par
%
%---------------------------------------------------------------
% 
In section~2.3, we have provided a thorough study of the corresponding Bargmann wave
functions, which in contrast with what is claimed in~\cite{loutsenko} does not happen to
be very simple. It indeed turns out that whilst the ground and first-excited states (see
(\ref{eq:gs}), (\ref{eq:excited}) and (\ref{eq:P1})) are correctly given by equations (38)
and (39) of that reference after identifying the parameters as above-mentioned and $z$
with $\xi$, such is not the case for the higher-excited states, which as immediately seen from
(\ref{eq:P2}) and (\ref{eq:P3}), do not factorize as stated in~\cite{loutsenko}.
Furthermore, our approach has directly led us to a compact expression for the
ground-state wave function (\ref{eq:gs}) as a product of two $q$-exponentials instead of
the infinite product of quadratic factors displayed in~\cite{loutsenko}. The former
can be easily expanded into the latter by employing well-known properties of
$q$-exponentials~\cite{ubriaco, cq03c}.\par 
%
%======================================================
% 
\section{\boldmath $D$-dimensional harmonic oscillator}
\setcounter{equation}{0}

Let us now consider the $D$-dimensional harmonic oscillator problem described by the
Hamiltonian
\begin{equation}
  H = \frac{\pb^2}{2m} + \frac{1}{2} m \omega^2 \xb^2  \label{eq:D-H}
\end{equation}
where the position and momentum components $x_i$, $p_i$, $i=1$, 2,~\ldots, $D$, satisfy
modified commutation relations of the type~\cite{kempf97, kempf95, chang}
\begin{eqnarray}
  [x_i, p_j] & = & {\rm i}\hbar \left(\delta_{i,j} + \bbeta \pb^2 \delta_{i,j} + \bbeta' p_i p_j
       \right) \nonumber \\{}
  [p_i, p_j] & = & 0  \label{eq:D-com} \\{}
  [x_i, x_j] & = & {\rm i}\hbar \frac{(2\bbeta - \bbeta') + (2\bbeta + \bbeta') \bbeta \pb^2}
       {1 + \bbeta \pb^2} (p_i x_j - p_j x_i) \nonumber 
\end{eqnarray}
where $\bbeta$, $\bbeta'\ge 0$. Such commutation relations imply isotropic nonzero
minimal uncertainties in the position coordinates $\Delta x_{0i} = \Delta x_0 = \hbar \sqrt{D
\bbeta + \bbeta'}$, but none in the momentum coordinates, which are simultaneously
diagonalizable.\par
%
%---------------------------------------------------------------------------------------
%
In the momentum representation, the operators $p_i$ become multiplicative operators, while
the operators $x_i$ are realized as differential operators ${\rm i}\hbar \left[(1 + \bbeta p^2)
\partial/\partial p_i + \bbeta' p_i p_j \partial/\partial p_j + \ggamma p_i\right]$. Here
$\ggamma$ is an arbitrary constant, which does not appear in the commutation relations
(\ref{eq:D-com}) and only affects the weight function in the scalar product in momentum
space
\begin{equation}
  \langle f | g \rangle = \int \frac{d^D \pb}{\left[1 + (\bbeta + \bbeta')
  p^2\right]^{1-\alpha}} f^*(\pb) g(\pb)  \label{eq:scal-prod}
\end{equation}
where
\begin{equation}
  \alpha = \frac{\ggamma - \frac{1}{2} (D-1) \bbeta'}{\bbeta + \bbeta'}.
\end{equation}
\par
%
%----------------------------------------------------------------------------------------
%
Since the Hamiltonian (\ref{eq:D-H}) is rotationally invariant, its momentum space
eigenfunctions can be expressed as a product of a D-dimensional spherical harmonics
$Y_{l_{D-1}\cdots l_2 l_1}(\Omega)$ and a radial wave function $R_{nl}(p)$ (where $l =
l_{D-1}$). In terms of a dimensionless variable $P = pa/\hbar$ and dimensionless parameters
$\beta = \bbeta \hbar^2/a^2$, $\beta' = \bbeta' \hbar^2/a^2$, $\gamma = \ggamma
\hbar^2/a^2$, where $a = \sqrt{\hbar/(m\omega)}$, the radial differential
equation~\cite{chang} can be written as
\begin{eqnarray}
  &&\frac{1}{2} \Biggl\{- \Biggl(f(P) \frac{d}{dP}\Biggr)^2 - \Biggl[\frac{D-1}{P} +
       \Bigl((D-1) \beta + 2\gamma\Bigr)P\Biggr] f(P) \frac{d}{dP} + \frac{L^2}{P^2} -
       (D \gamma - 2\beta L^2) \nonumber \\
  && \mbox{} + \Bigl[1 + \beta^2 L^2 - \gamma (D \beta + \beta' + \gamma)\Bigr] P^2
       \Biggr\} R_{nl}(P) = e_{nl} R_{nl}(P)  \label{eq:radial-eqn} 
\end{eqnarray}   
where $L^2$ is the eigenvalue of the square of the $D$-dimensional angular momentum
\begin{equation}
  L^2 = l(l + D -2) \qquad l=0, 1, 2, \ldots
\end{equation}
$n$ is the radial quantum number and
\begin{equation}
  f(P) = 1 + \beta_0 P^2 \qquad \beta_0 = \beta + \beta' \qquad e_{nl} = \frac{E_{nl}}
  {\hbar\omega}.
\end{equation}
\par
%
%-------------------------------------------------------------------------------------
%
The first-order derivative in (\ref{eq:radial-eqn}) can be eliminated by setting
\begin{equation}
  R_{nl}(P) = P^{-(D-1)/2} [f(P)]^{-\alpha/2} \chi_{nl}(P).  \label{eq:radial-fct}
\end{equation}
The resulting equation reads
\begin{equation}
  h^{(l)} \chi_{nl}(P) = \te_{nl} \chi_{nl}(P)  \label{eq:radial-eqn-bis}
\end{equation}
where
\begin{eqnarray}
  h^{(l)} & = & \frac{1}{2}\left\{- \left[f(P) \frac{d}{dP}\right]^2 + \frac{a^{(l)}}{P^2} +
       b^{(l)} P^2\right\}  \label{eq:hl} \\
  a^{(l)} & = & L^2 + \frac{1}{4} (D-3)(D-1) = \left(l + \frac{D-3}{2}\right) \left(l +
       \frac{D-1}{2}\right) \\
  b^{(l)} & = & 1 + \beta^2 \left[L^2 + \frac{1}{4} (D^2-1)\right] + \frac{1}{2} (D-1) \beta
       \beta' \\
  \te_{nl} & = & e_{nl} - \beta \left[L^2 + \frac{1}{4} (D-1)^2\right] + \frac{1}{4} (D-1)
       \beta'.  \label{eq:t-e} 
\end{eqnarray}
From (\ref{eq:radial-fct}), it follows that in the space spanned by functions $\chi(P)$, the
scalar product (\ref{eq:scal-prod}) becomes
\begin{equation}
  \langle\chi' | \chi\rangle = \int_0^\infty \frac{dP}{f(P)} \chi^{\prime*}(P) \chi(P).
  \label{eq:scal-prod-bis}
\end{equation}
One should note that the arbitrary constant $\ggamma$, appearing in the momentum space
realization of $x_i$ and in the scalar product (\ref{eq:scal-prod}), is absent from equations
(\ref{eq:radial-eqn-bis}) -- (\ref{eq:scal-prod-bis}). From the very beginning, it is therefore
obvious that neither $\te_{nl}$ (hence $e_{nl}$) nor $\chi_{nl}(P)$ can depend on it.\par
%
%++++++++++++++++++++++++++++++++++++++++++++++++++
%
\subsection{Energy spectrum}

Let us first prove that $h^{(l)}$ can be factorized as
\begin{equation}
  h^{(l)} = B^+(g,s) B^-(g,s) + \tepsilon_0  \label{eq:D-h} 
\end{equation}
where
\begin{equation}
  B^{\pm}(g,s) = \frac{1}{\sqrt{2}} \left(\mp f(P) \frac{d}{dP} + gP - \frac{s}{P}\right)
  \label{eq:D-B+/-}
\end{equation}
and $\tepsilon_0$ is the factorization energy. Here $g$ and $s$ are assumed to be two
positive constants that are functions of $l$ and of the parameters $\beta$, $\beta'$ of the
problem.\par
%
%--------------------------------------------------------------------------------------
%
On inserting (\ref{eq:D-B+/-}) in (\ref{eq:D-h}) and comparing the result with
(\ref{eq:hl}), we get the three conditions
\begin{eqnarray}
  s(s-1) & = & a^{(l)} \\
  g(g-\beta_0) & = & b^{(l)} \\
  \tepsilon_0 & = & gs + \frac{1}{2} (g + \beta_0 s).  \label{eq:epsilon-0}
\end{eqnarray}
Their solution is given by
\begin{eqnarray}
  s & = & l + \frac{1}{2}(D-1)  \label{eq:s} \\
  g & = & \frac{1}{2} \beta_0 + \Delta^{(l)} \qquad \Delta^{(l)} = \sqrt{1 +
        \beta^2 L^2 + \frac{1}{4}(D \beta + \beta')^2}  \label{eq:g} \\
  \tepsilon_0 & = & \frac{1}{2} \beta_0 \left(2l + D - \frac{1}{2}\right) + \left(l +
        \frac{D}{2}\right) \Delta^{(l)}.
\end{eqnarray}
In the limit $\beta$, $\beta' \to 0$, we get $s = l + \frac{1}{2}(D-1)$, $g \to 1$,
$\tepsilon_0 \to l + \frac{D}{2}$, which correspond to the usual factorization for
the $D$-dimensional harmonic oscillator in conventional quantum mechanics (see,
e.g.,~\cite{fernandez} for the three-dimensional case).\par
%
%---------------------------------------------------------------
%  
The next step consists in considering a hierarchy of Hamiltonians
\begin{equation}
  h^{(l)}_i = B^+(g_i, s_i) B^-(g_i, s_i) + \sum_{j=0}^i \tepsilon_j \qquad  i=0,
  1, 2, \ldots  \label{eq:D-hierarchy} 
\end{equation}
where $h^{(l)}_0 = h^{(l)}$, $g_0 = g$, $s_0 = s$ and $g_i$, $s_i$, $i=1$, 2,~\ldots,
are some positive constants. On imposing a shape invariance condition similar to
equation (\ref{eq:SI}), we obtain the set of three relations
\begin{eqnarray}
  s_{i+1}(s_{i+1} - 1) & = & s_i(s_i + 1) \\
  g_{i+1}(g_{i+1} - \beta_0) & = & g_i(g_i + \beta_0) \\
  \tepsilon_{i+1} & = & g_{i+1}\left(s_{i+1} + \frac{1}{2}\right) - g_i\left(s_i -
       \frac{1}{2}\right) + \frac{1}{2}\beta_0 (s_{i+1} + s_i).
\end{eqnarray}
\par
%
%----------------------------------------------------------------
% 
The solution of the first two is given by
\begin{equation}
  s_i = s + i \qquad g_i = g + \beta_0 i.  \label{eq:si}
\end{equation}
The third one, together with equation (\ref{eq:epsilon-0}), leads to the eigenvalues
\begin{equation}
  \te_n(g,s) = \sum_{i=0}^n \tepsilon_i = g\left(2n + s + \frac{1}{2}\right) +
  \beta_0 s\left(2n + \frac{1}{2}\right) + 2\beta_0 n^2.
\end{equation}
\par
%
%-----------------------------------------------------------------
%
Taking equations (\ref{eq:s}) and (\ref{eq:g}) into account, we can rewrite them as
\begin{equation}
  \te_{nl} = \left(2n + l + \frac{D}{2}\right) \Delta^{(l)} + \frac{1}{2}\beta_0
  \left(2n + l + \frac{D}{2}\right) + \beta_0 \left(l + \frac{D-1}{2}\right)\left(2n +
  \frac{1}{2}\right) + 2\beta_0 n^2  \label{eq:te-nl}
\end{equation}
or 
\begin{equation}
  \te_{Nl} = \left(N + \frac{D}{2}\right) \Delta^{(l)} + \frac{1}{2}\beta_0
  \left(N^2 +DN - L^2 + D - \frac{1}{2}\right)  \label{eq:te-Nl} 
\end{equation}
in terms of $l$ and of either the radial quantum number $n$ or the principal quantum
number $N = 2n+l$. Finally, from equation (\ref{eq:t-e}), it follows that the eigenvalues in
the radial differential equation (\ref{eq:radial-eqn}) can be expressed as
\begin{equation}
  e_{Nl} = \left(N + \frac{D}{2}\right) \Delta^{(l)} + \frac{1}{2}\left[(\beta + \beta')
  \left(N + \frac{D}{2}\right)^2 + (\beta - \beta')\left(L^2 + \frac{D^2}{4}\right) +
  \beta' \frac{D}{2}\right].  \label{eq:eigenvalue}
\end{equation}
\par
%
%-------------------------------------------------------------------------------------------
%
In the limit $\beta$, $\beta' \to 0$, we obtain that $\te_{Nl}$ and $e_{Nl}$ go to $e_N
= N + \frac{D}{2}$, which is the conventional result. Equation (\ref{eq:eigenvalue}) shows
that for nonvanishing $\beta$, $\beta'$, the spectrum of $H$, given by $E_{Nl} = \hbar
\omega e_{Nl}$, is quadratic in $N$ with an additional $l$ dependence absent in the
conventional case. The values obtained for $E_{Nl}$ coincide with those given in equation
(57) of~\cite{chang}.\par
%
%---------------------------------------------------------------------------------------
%
The results obtained for $g_i$, $s_i$ and $\tepsilon_i$ allow us to write the Hamiltonians
(\ref{eq:D-hierarchy}) of the SUSYQM hierarchy as
\begin{equation}
  h^{(l)}_i = \frac{1}{2}\left\{- \left[f(P) \frac{d}{dP}\right]^2 + \frac{a^{(l)}_i}{P^2} +
       b^{(l)}_i P^2\right\} + c^{(l)}_i  \label{eq:D-h-i}
\end{equation}
where
\begin{eqnarray}
  a^{(l)}_i & = & \left(l + i + \frac{D-3}{2}\right) \left(l + i + \frac{D-1}{2}\right) \\
  b^{(l)}_i & = & b^{(l)} + 2\beta_0 i \Delta^{(l)} + \beta_0^2 i^2 \\
  c^{(l)}_i & = & i \left[\beta_0\left(l + i - 1 + \frac{D}{2}\right) + \Delta^{(l)}\right].
\end{eqnarray}
In the limit $\beta$, $\beta' \to 0$, $a^{(l)}_i$ remains unchanged while $b^{(l)}_i \to
1$ and $c^{(l)}_i \to i$.\par
%
%--------------------------------------------------------------------------------------------
%
As in conventional quantum mechanics, the supersymmetric partners coincide formally
with some radial Hamiltonians corresponding to shifted $l$ values, $l+i$, where $i=1$,
2,~\ldots. It should be stressed that the angular part $Y_{l_{D-1} \cdots l_2
l_1}(\Omega)$ of the wave functions being left unchanged, supersymmetry only
concerns here the radial equation.\par
%
%++++++++++++++++++++++++++++++++++++++++++++++++++++++
%
\subsection{Radial wave functions}

In this subsection, we plan to determine the explicit form of the eigenfunctions
$\chi_{nl}(P)$ of $h^{(l)}$. Since $(h^{(l)})^{\dagger} = h^{(l)}$ with respect to scalar
product (\ref{eq:scal-prod-bis}), such eigenfunctions satisfy the orthonormality relation
\begin{equation}
  \int_0^{\infty} \frac{dP}{f(P)} \chi^*_{n'l}(P) \chi_{nl}(P) = \delta_{n',n}.
\end{equation}
Note that with respect to (\ref{eq:scal-prod-bis}), we also have $(B^+(g,s))^{\dagger} =
B^-(g,s)$.\par
%
%------------------------------------------------------------------------------------------
%
The ground state wave function $\chi_{0l}(P) = \chi_0(g,s;P)$ of $h^{(l)}$ is obtained
from the condition
\begin{equation}
  B^-(g,s) \chi_0(g,s;P) = 0
\end{equation}
and given by
\begin{equation}
  \chi_{0}(g,s;P) = {\cal N}_0(g,s) P^s [f(P)]^{-(g+\beta_0 s)/(2\beta_0)} = {\cal
N}_{0}(g,s)
  P^{\mu + \frac{1}{2}} [f(P)]^{-\frac{1}{2}(\lambda+\mu+1)}. 
\end{equation}
Here $s$ and $g$ are given by equations (\ref{eq:s}) and (\ref{eq:g}), respectively,
$\lambda$ and $\mu$ are defined by
\begin{equation}
 \lambda = \frac{1}{\beta_0}\left(g - \frac{1}{2}\beta_0\right) \qquad \mu = s - 
  \frac{1}{2}  \label{eq:D-lambda}
\end{equation}
and the normalization coefficient is
\begin{equation}
  {\cal N}_{0l} = {\cal N}_0(g,s) = \left(\frac{2 \Gamma(\lambda+\mu+2)} 
  {\Gamma(\lambda+1) \Gamma(\mu+1)} \beta_0^{\mu+1}\right)^{1/2}.
\end{equation}
\par
%
%-------------------------------------------------------------------------------------
%
The excited state wave functions $\chi_{n,l}(P) = \chi_n(g,s;P)$, $n=1$, 2,~\ldots, can
be obtained from the recursion relation
\begin{eqnarray}
  \chi_{n+1}(g,s;P) & = & [\beta_0 (n+1) (n+\lambda+\mu+2)]^{-1/2} \frac{1}{2}
  \left[- f(P) \frac{d}{dP} + \beta_0 \left(\lambda + \frac{1}{2}\right) P - \frac{\mu +
  \frac{1}{2}}{P}\right] \nonumber \\
  && \mbox{} \times \chi_n(g_1,s_1;P) \qquad n=0, 1, 2, \ldots  \label{eq:D-recursion}  
\end{eqnarray}
where from (\ref{eq:si}) and (\ref{eq:D-lambda}), it follows that $g_1 = g+\beta_0$,
$s_1 = s+1$ correspond to $\lambda_1 = \lambda+1$, $\mu_1 = \mu+1$.\par
%
%-------------------------------------------------------------------------------------------
%
Let us set
\begin{equation}
  \chi_n(g,s;P) = {\cal N}_n(g,s) P_n^{(\lambda, \mu)}(z) P^{\mu + \frac{1}{2}}
  [f(P)]^{-\frac{1}{2}(\lambda+\mu+1)}  \label{eq:chi-n} 
\end{equation}
where ${\cal N}_{nl} = {\cal N}_n(g,s)$ is some normalization coefficient and
$P_n^{(\lambda, \mu)}(z)$ is some $(\lambda,\mu)$-dependent, $n$th-degree
polynomial in the variable
\begin{equation}
  z = \frac{\beta_0 P^2 - 1}{1 + \beta_0 P^2}
\end{equation}
varying in the range $(-1, +1)$ (with $P_0^{(\lambda, \mu)}(z) \equiv 1$). Inserting
(\ref{eq:chi-n}) in (\ref{eq:D-recursion}) converts the latter into the relation
\begin{eqnarray}
  P^{(\lambda,\mu)}_{n+1}(z) & = & [\beta_0 (n+1) (n+\lambda+\mu+2)]^{-1/2}
        \frac{{\cal N}_n(g_1,s_1)}{{\cal N}_{n+1}(g,s)} \nonumber \\ 
  && \mbox{} \times \frac{1}{2} \left[- (1 - z^2) \frac{d}{dz} + \lambda - \mu + (\lambda +
        \mu + 2) z\right] P^{(\lambda+1,\mu+1)}_n(z). \label{eq:backward} 
\end{eqnarray} 
\par
%
%---------------------------------------------------------------------------------------
% 
The differential operator on the right-hand side of (\ref{eq:backward}) can be recognized
as the backward shift operator for Jacobi polynomials, satisfying the
property~\cite{koekoek}
\begin{equation}
  \left[- (1 - z^2) \frac{d}{dz} + \lambda - \mu + (\lambda + \mu + 2) z\right]
  P^{(\lambda+1,\mu+1)}_n(z) = 2(n+1) P^{(\lambda,\mu)}_{n+1}(z).
\end{equation}
Hence the polynomials of equation (\ref{eq:chi-n}) can be identified with Jacobi
polynomials, while ${\cal N}_n(g,s)$ satisfies the recursion relation
\begin{equation}
  {\cal N}_{n+1}(g,s) = \left(\frac{n+1}{\beta_0(n+\lambda+\mu+2)}\right)^{1/2}
  {\cal N}_n(g_1,s_1)
\end{equation}
from which we get
\begin{equation}
  {\cal N}_n(g,s) = \left(\frac{2 (2n+\lambda+\mu+1) n!\, \Gamma(n+\lambda+\mu+1)}
  {\Gamma(n+\lambda+1) \Gamma(n+\mu+1)} \beta_0^{\mu+1}\right)^{1/2}.
  \label{eq:N-n}
\end{equation}
\par
When combining equations (\ref{eq:chi-n}) and (\ref{eq:N-n}) with (\ref{eq:radial-fct}),
we finally obtain the same result as in equation (58) of~\cite{chang}.\par
%
%++++++++++++++++++++++++++++++++++++++++++++++++++++
%
\subsection{Alternative factorizations}

As a final point, we would like to comment on the factorization of $h^{(l)}$ carried out in
section~3.1.\par
%
%------------------------------------------------------------------------------------------
%
Equation (\ref{eq:D-hierarchy}) and the corresponding shape invariance condition show
that each element of the hierarchy $h^{(l)}_i$, $i=0$, 1, 2,~\ldots, admits two different
factorizations: this is an example of the so-called two-way factorization~\cite{delange}.
The first-order operators $B^{\pm}(g_i,s_i)$ involved in the factorization act as shift
operators, connecting pairs of eigenstates with the same energy, belonging to two
consecutive Hamiltonians of the hierarchy.\par
%
%-----------------------------------------------------------------------------------------
%
The conventional harmonic oscillator radial equation is known to admit a four-way
factorization~\cite{fernandez}: there indeed exists another pair  of factorizations
associated with another hierarchy of Hamiltonians $h^{(l)\prime}_i$, $i=0$, 1, 2,~\ldots.
As before, the corresponding first-order operators $B^{\pm\prime}(g_i,s_i)$ connect
pairs of eigenstates with the same energy, belonging to two consecutive Hamiltonians of
the second hierarchy. Since, however, the Hamiltonians of the two extended
hierarchies\footnote{By extended hierarchies, we mean the sets of Hamiltonians
$h^{(l)}_i$ or $h^{(l)\prime}_i$ obtained by letting $i$ run over  {\bf Z}. Some of these
Hamiltonians may be unphysical. For instance, in the case of $h^{(l)}_i$ given in
(\ref{eq:D-h-i}), unphysical Hamiltonians correspond to $i < -l$.} are linked through the
relation $h^{(l)\prime}_{-i} = h^{(l)}_i - 2i$, the shift operators of the second hierarchy
also connect pairs of eigenstates with different energies, belonging to two consecutive
Hamiltonians of the first hierarchy. For such a reason, one can combine both pairs of shift
operators to construct ladder operators connecting the eigenstates of the same
Hamiltonian~\cite{fernandez} (see also~\cite{delsol}).\par
%
%-----------------------------------------------------------------------------------
% 
In the case of the deformed commutation relations (\ref{eq:D-com}), the harmonic
oscillator radial equation (\ref{eq:radial-eqn-bis}) also admits a pair of alternative
factorizations, where in the counterpart of (\ref{eq:D-B+/-}) we choose $g>0$ and $s<0$
instead of both $g$, $s>0$. Distinguishing by primes all parameters relative to these
alternative factorizations from those of section~3.1, we obtain
\begin{eqnarray}
  s' & = & - l - \frac{1}{2}(D-3) \\
  g' & = & \frac{1}{2} \beta_0 + \Delta^{(l)} \\
  \tepsilon_0' & = & - \frac{1}{2} \beta_0 \left(2l + D - \frac{7}{2}\right) - \left(l +
        \frac{D-4}{2}\right) \Delta^{(l)}
\end{eqnarray}
and
\begin{eqnarray}
  s'_i & = & s' + i \\
  g'_i & = & g' + \beta_0 i \\
  \tepsilon'_{i} & = & g'_i\left(s'_i + \frac{1}{2}\right) - g'_{i-1}\left(s'_{i-1} -
       \frac{1}{2}\right) + \frac{1}{2}\beta_0 (s'_i + s'_{i-1})
\end{eqnarray}
for  $i=1$, 2,~\ldots. For $\beta$, $\beta' \to 0$, we get $s' = - l - \frac{1}{2}(D-3)$,
$g' \to 1$, $\tepsilon'_0 \to - l - \frac{1}{2}(D-4)$ and $s'_i = - l + i - \frac{1}{2}(D-3)$,
$g'_i \to 1$, $\tepsilon'_i \to 2$, which agree with the conventional
results~\cite{fernandez}.\par
%
%-------------------------------------------------------------------------------------------------
%
Proceeding as in section~3.1, we can write the energy eigenvalues $\te_m(g',s') =
\sum_{i=0}^m \tepsilon'_i$ of $h^{(l)}$ as 
\begin{eqnarray}
  \te_{ml} & = & \left(2m - l - \frac{D-4}{2}\right) \Delta^{(l)} + \frac{1}{2}\beta_0
  \left(2m - l - \frac{D-4}{2}\right) - \beta_0 \left(l + \frac{D-3}{2}\right)\left(2m +
  \frac{1}{2}\right) \nonumber \\
  && \mbox{} + 2\beta_0 m^2.
\end{eqnarray}
The comparison with equations (\ref{eq:te-nl}) and (\ref{eq:te-Nl}) shows that the new
quantum number $m$ is related to the radial and principal quantum numbers, $n$ and
$N$, through the relation
\begin{equation}
  m = n + l + \frac{1}{2}(D-2) = \frac{1}{2}(N + l + D - 2).
\end{equation}
We have therefore rederived the energy spectrum of $h^{(l)}$ in an alternative way.\par
%
%------------------------------------------------------------------------------------------
%
The second Hamiltonian hierarchy $h^{(l)\prime}_i$, $i=0$, 1, 2,~\ldots, containing
(\ref{eq:hl}) as its first member, is given by an expression similar to (\ref{eq:D-h-i}) with
\begin{eqnarray}
  a^{(l)\prime}_i & = & \left(l - i + \frac{D-3}{2}\right) \left(l - i + \frac{D-1}{2}\right) \\
  b^{(l)\prime}_i & = & b^{(l)} + 2\beta_0 i \Delta^{(l)} + \beta_0^2 i^2 \\
  c^{(l)\prime}_i & = & i \left[\beta_0\left(- l + i + 1 - \frac{D}{2}\right) +
\Delta^{(l)}\right].
\end{eqnarray}
For $\beta$, $\beta' \to 0$, $a^{(l)\prime}_i$ remains unchanged, but $b^{(l)\prime}_i
\to 1$ and $c^{(l)\prime}_i \to i$, so that we recover the property $h^{(l)\prime}_{-i} =
h^{(l)}_i - 2i$ for the Hamiltonians of the extended hierarchies. For nonvanishing $\beta$,
$\beta'$, however, there is no simple relation between them and the only common
member is $h^{(l)} = h^{(l)}_0 = h^{(l)\prime}_0$. Ladder operators cannot therefore be
built by combining the two different types of shift operators. Such ladder operators have
actually been constructed by using another method~\cite{dadic} and they appear to be
very complicated operators.\par
%
%============================================================
%
\section{Conclusion}

In the present paper, we have continued with the application, initiated in I, of combined
factorization and shape invariance techniques to quantum mechanical problems in the context
of a theory based on some deformed canonical commutation relations and predicting
nonzero minimal uncertainties in position and/or momentum.\par
%
%----------------------------------------------------------------------------------------
%
To start with, we have determined for the first time the spectrum and the eigenvectors of
a one-dimensional harmonic oscillator in a uniform electric field $\e$. We have established
that whenever $\alpha \ne 0$, i.e., whenever there is a nonzero minimal uncertainty in
momentum, the correction to the harmonic oscillator eigenvalues due to the electric field
depends on the energy level and actually decreases when going up in energy. This is true
whether there is also a nonzero minimal uncertainty in position or no, i.e., whether $\beta
\ne 0$ or $\beta=0$. As was shown in I, in the former case the harmonic oscillator
spectrum is exponential, whereas in the latter it is quadratic. In contrast, whenever
$\alpha=0$ and $\beta \ne 0$, the electric field induces a constant shift of the
(quadratic) harmonic oscillator spectrum, as is the case in conventional quantum
mechanics.\par
%
%----------------------------------------------------------------------------------------
%
In addition, we have shown that although the electric field remains the same in all the
supersymmetric partners of our original Hamiltonian, its effect on the eigenvectors is
rather complicated. The latter have been fully determined for $\alpha = \beta \ne 0$, in
which case the harmonic oscillator with nonzero minimal uncertainties in both position and
momentum reduces to a $q$-deformed harmonic oscillator with $q>1$. In the general
case $0 \ne \alpha \ne \beta \ne 0$, the ground state and the first few excited states
have been given in terms of $n$-$q$-boson states. In both instances, use has been made
of a ($q$-deformed) Bargmann representation of the latter and of $q$-differential
calculus.\par
%
%-----------------------------------------------------------------------------------------
%
Furthermore, we have shown how our results compare with those of a previous study of
the spectrum of the most general Hermitian Hamiltonian that is bilinear in $q$-boson
creation and annihilation operators with $q>1$.\par
%
%-------------------------------------------------------------------
%
In the second part of our paper, we have reconsidered the problem of a $D$-dimensional
harmonic oscillator when there are isotropic nonzero minimal uncertainties in the position
coordinates, depending on two parameters $\beta$, $\beta'$. We have established that
our SUSYQM techniques can be extended to deal with the corresponding radial equation in the
momentum representation. As a result, we have rederived in a very simple way both the
spectrum and the momentum radial wave functions, previously found through lengthy
differential equation techniques~\cite{chang}. Finally, we have commented on various
factorizations of the radial Hamiltonian and stressed both the resemblances and the differences
between the deformed case and the conventional one.\par
%
%----------------------------------------------------------------------------------------------
%
The second part of our paper opens the way to other yet unsolved $D$-dimensional
problems. Among these, we may quote that of a $D$-dimensional harmonic oscillator in a
magnetic field, which would be in line with the first part of our paper.\par
%
%=====================================================
%
\section*{Acknowledgments}

CQ is a Research Director of the National Fund for Scientific Research (FNRS), Belgium.
VMT is very grateful to Professor C.\ Quesne for warm hospitality at Universit\'e Libre de
Bruxelles and thanks the National Fund for Scientific Research (FNRS), Belgium, for
financial support.\par
%
%======================================================
% 

\end{document}